\documentstyle[12pt,a41,epsfig]{article}
\newcommand{\MeV}{{\rm MeV}}
\newcommand{\GeV}{{\rm GeV}}

\newcommand{\lsim}{\raisebox{-0.07cm   } 
{$\, \stackrel{<}{{\scriptstyle\sim}}\, $}}

\begin{document}
\sloppy
\thispagestyle{empty}
\begin{flushleft}
DESY 03--049    \hfill
{\tt hep-ph/0304292}\\
INLO-PUB-07/03\\
SFB/CPP--03--08\\
April 2003  \\
\end{flushleft}

\mbox{}
\vspace*{\fill}
\begin{center}
 
{\LARGE\bf Twist--2 Heavy Flavor Contributions to the }

\vspace{2mm}
{\LARGE\bf
Structure
Function {\boldmath $g_2(x,Q^2)$}}

\vspace{4em}
\large
J. Bl\"umlein$^a$, V. Ravindran$^{a,b}$ and W.L. van Neerven$^{a,c}$
\normalsize
\\
\vspace{4em}
{\it$^a$DESY Zeuthen}\\
{\it    Platanenallee 6, D--15738 Zeuthen, Germany}\\

\vspace{4mm}
{\it $^b$Harish--Chandra
Research Institute}\\
{\it Chhatnag Road, Jhusi, Allahabad, 211019, India}\\

\vspace{4mm}
{\it 
$^c$Instituut-Lorentz,
Universiteit Leiden,}\\ {\it
P.O. Box 9506, 2300 HA Leiden, The Netherlands.}
\end{center}
\vspace*{\fill}
\begin{abstract}
\noindent
The twist--2 heavy flavor contributions to the polarized structure function 
$g_2(x,Q^2)$ are calculated. We show that this part of $g_2(x,Q^2)$ is related 
to the heavy flavor contribution to $g_1(x,Q^2)$ by the Wandzura--Wilczek 
relation to all orders in the strong coupling constant. Numerical results 
are presented.
\end{abstract}
\vspace*{\fill}
\newpage
\section{Introduction}

\vspace{1mm}
\noindent
Heavy flavor contributions to polarized and unpolarized deep inelastic 
structure functions due to charm and bottom--quark production cause 
different scaling violations if compared to those due to massless partons. 
Moreover it is known that in certain phase space regions of $x$ and $Q^2$ 
these contributions can be large~\cite{UNPLA}. Reliable determinations of
the QCD scale $\Lambda_{\rm QCD}$ therefore require a careful account of
the heavy flavor contributions. In the case of the unpolarized structure
functions $F_{2,L}(x,Q^2)$ and $xF_3(x,Q^2)$ the twist--2 contributions 
with $Q\overline{Q}$--final states, $Q = c, b$ were calculated to 
next--to--leading order (NLO)~\cite{UNP,BVN}. For polarized 
deep--inelastic
scattering only the leading order corrections are known for the structure
function $g_1(x,Q^2)$ at present~\cite{POLLO}. 
All corrections mentioned above have been calculated using mass 
factorization. As shown in~[5--8]
this
method fails in the case of the polarized structure functions which
emerge for transverse polarization already in the massless quark limit.
The reason for this lies in a violation of covariance due to the
omission of contributions $\propto S.k$, where $S$ denotes the
spin vector of the nucleon and $k$ the parton 4--momentum.

In Refs.~\cite{RR,BK1} 
it was shown that the quarkonic contributions to
the transverse structure functions can be correctly obtained using the
covariant parton model. The results in this approach are identical to
those derived in the local light--cone expansion, cf.~\cite{WW,BK2},
for massless quarks on the level of the twist--2 contributions. In
particular the twist--2 part of $g_2(x,Q^2)$ is obtained by the
Wandzura--Wilczek relation. The effect of quark masses for quarkonic
matrix elements was further investigated in Ref.~\cite{BT} using the 
method of \cite{GP}. Also in this case the Wandzura--Wilczek relation
was found to hold, irrespective of the values of the quark masses chosen.
In \cite{RID,BT} the target mass corrections to the quarkonic 
contributions to $g_2(x,Q^2)$ and the other four polarized structure
functions~\cite{BT} were studied. Again the Wandzura--Wilczek relation
was found to hold relating the twist--2 contributions of $g_1(x,Q^2)$
and $g_2(x,Q^2)$. In a more general approach the amplitudes which
contribute to the quarkonic matrix elements for deeply virtual 
non--forward scattering were investigated in \cite{BGR}. In the generalized
Bjorken limit the non--forward Compton amplitude is expressed by
operator matrix elements of vector operators. However, one usually
parameterizes it in terms of scalar operator matrix elements to which
the (generalized) parton distributions correspond. Therefore relations
between the operator matrix elements of the vector and scalar operators
are implied, which form the origin of the Wandzura--Wilczek relations
and other integral--relations~\cite{BK2,BT}. The absence of the typical
integral terms e.g. in the Callan--Gross~\cite{CG} relation is merely
the exception and caused by a cancellation of the former,~see 
Ref.~\cite{BR1}. For the polarized case it was shown that the
Wandzura--Wilczek relation holds for the twist--2 contributions to
the respective non--forward amplitudes, cf.~\cite{BR1}. This result could 
be generalized allowing multiple meson production in the final state 
in~\cite{BEGR}. Also 
semi--inclusive processes such as diffrative scattering have been 
investigated w.r.t. the emergence of Callan--Gross and Wandzura--Wilczek 
relations~\cite{BR2}. Although the variables change, the Wandzura--Wilczek
relation relates the twist--2 contributions of 
$g_1^{\rm diffr}(x,x_P,Q^2)$ to $g_2^{\rm diffr}(x,x_P,Q^2)$ independently
of the value for $x_P$. Integral relations of similar kind were also 
established for other structure functions and matrix 
elements~\cite{GL} and for the fermionic twist--3 contributions 
in~\cite{BT}.

In the present paper we calculate the heavy flavor twist--2 contribution
to the structure function $g_2(x,Q^2)$. We 
show that the Wandzura--Wilczek relation holds also in this case and 
present numerical predictions.

\section{Longitudinal Gluon Polarization}

\vspace{1mm} 
\noindent 
The hadronic tensor of the polarized part of the
$eN$--scattering cross section in case of pure photon
exchange\footnote{The corresponding expressions in the case of additional
weak boson exchange are given in \cite{BT}.} reads
\begin{eqnarray} 
\label{eqHAD}
W^{(A)}_{\mu\nu} = i\varepsilon_{\mu\nu\lambda\sigma}
\frac{q^\lambda S^\sigma}{p.q} g_1(x,Q^2)
+i\varepsilon_{\mu\nu\lambda\sigma} \frac{q^\lambda (p.q
S^\sigma-S.q p^\sigma)}{(p.q)^2} g_2(x,Q^2)~. 
\label{had1}
\end{eqnarray}
$S_\sigma$ denotes the nucleon spin vector, $p$ the nucleon momentum, and 
$q$ the vector of the 4--momentum transfer, with $Q^2 = -q^2$ and
$x= Q^2/(2 p.q)$.
The polarized part of the scattering cross section for longitudinal
nucleon polarization $S_L$, integrated over the azimuthal angle $\phi$,
is
\begin{eqnarray} \frac{d^2\sigma(\lambda, \pm S_L)}{dxdy} &=& \pm 2 
\pi{\sf S}
\frac{\alpha^2}{Q^4} \left [ -2\lambda y \left( 2-y-\frac{2 x y 
M^2}{\sf S}
\right) xg_1(x,Q^2) + 8 \lambda \frac{y x^2 M^2}{\sf S} g_2(x,Q^2) 
\right]~.  
\label{scaL} \end{eqnarray}
Correspondingly, for transversely polarized nucleons one obtains
\begin{eqnarray} \frac{d^3\sigma(\lambda, \pm S_T)} {dx dy d\phi} &=& \pm
{\sf S} \frac{\alpha^2}{Q^4}
 2 \sqrt{\frac{M^2}{\sf S}} \sqrt{x y \left [1 - y - \frac{x y M^2}{\sf 
S}
\right]} \cos(\chi-\phi)\nonumber\\ & &~~~~~~~~~~~~~~~~~~~~ 
\times
\left [
-2\lambda y x g_1(x,Q^2) -
4\lambda x  g_2(x,Q^2) \right]~.
\label{scaT}
\end{eqnarray}
Here $M$ is the nucleon mass, $\sf S$ the cms energy, $\alpha$ the fine 
structure constant, $y = 2p.q/{\sf S}$, $\lambda$ is the degree of lepton 
polarization and $S_T$ the degree of hadronic transverse polarization, 
$\chi$ denotes the azimuthal angle associated with $S_T^\mu$, 
(\ref{steq}), and $g_1(x,Q^2)$ and 
$g_2(x,Q^2)$ are the polarized structure functions which contribute in 
this case. 

The heavy flavor contributions to the longitudinal and transverse 
differential scattering cross sections are obtained in calculating the 
corresponding contributions to the polarized structure functions 
$g_1(x,Q^2)$ and $g_2(x,Q^2)$.

Let us consider the sub--system hadronic tensor for
photon--interactions with on--shell initial state partons
\begin{eqnarray}
\label{wAsub}
w_{\mu\nu}^{(A)} =   \frac{i}{q.k} \varepsilon_{\mu\nu\rho\sigma}
\left\{
q^\rho s^\sigma g_1^{\rm parton}(z,Q^2) + \left(
q^\rho s^\sigma - \frac{s.q}{q.k} q^\rho k^\sigma \right)
g_2^{\rm parton}(z,Q^2) \right\}~.
\end{eqnarray}
In assuming that the gluon is longitudinally polarized, i.e. parallel
to the proton {\sf and}  parton 4--momentum
\begin{eqnarray}
\label{eqSPI}
s_{\mu} = \xi_1 S_{\mu} = \xi_2 k_{\mu}~,
\end{eqnarray}
with $k = z p$, 
the sub--system hadronic tensor of the polarization asymmetry
(\ref{wAsub}) 
receives  purely longitudinal contributions, since the term
$\propto g_2^{\rm parton}$ vanishes. Clearly (\ref{eqSPI}) is a special
model assumption which does not describe the general case being discussed 
in section~3. However, it is possible to derive in this approximation the 
correct expression for the coefficient functions 
$C_{g_1}^{Q\overline{Q}}$ at twist--2 which contributes to 
the structure function $g_1^{Q\overline{Q}}(x,Q^2)$. 
At leading order in $\alpha_s$ the structure function 
$g_1^{Q\overline{Q}}(x,Q^2)$ receives only gluonic 
contributions and is obtained as~\cite{POLLO}
\begin{eqnarray}
\label{eqg1}
g_1^{Q\overline{Q}}(x,Q^2) = 2 e_Q^2 \frac{\alpha_s(Q^2)}{2\pi}
\int_{a x}^1 \frac{dy}{y} C_{g_1}^{Q\overline{Q}}\left(
\frac{x}{y},M_Q^2,Q^2\right) \Delta G(y,Q^2)~,
\end{eqnarray}
with
\begin{eqnarray}
\label{eqWILS}
C_{g_1}^{Q\overline{Q}}(z,Q^2) = \frac{1}{2} \left[\beta (3-4z) - (1-2z)
\ln \left|\frac{1+\beta}{1-\beta}\right|\right]
\end{eqnarray}
and $\Delta G(x,Q^2)$ the polarized gluon distribution.
Here $a$ denotes the threshold $a=1+4M^2_Q/Q^2$, $e_Q$ the charge of the 
produced heavy quarks, and $\beta$ is the cms velocity
\begin{eqnarray}
\beta = \sqrt{1 - \frac{4M_Q^2}{Q^2} \frac{z}{1-z}}
\end{eqnarray}
of the final state quarks, with $\beta~\epsilon~\left[0, \sqrt{1-4M_Q^2 
x/[Q^2(1-x)]}~\right]$. 
Eq.~(\ref{eqg1}) defines the LO twist--2 contribution to 
$g_1^{Q\overline{Q}}(x,Q^2)$~\footnote{Note, that mass--scale effects
may 
introduce twist--3 contributions to the structure function $g_1(x,Q^2)$ 
as well, as has been shown for target mass corrections in 
Ref.~\cite{BT}.}. Note that 
\begin{eqnarray}
\label{fmom}
\int_0^{1/a} dz C_{g_1}^{Q\overline{Q}}(z,M_Q^2,Q^2) = 0~, 
\end{eqnarray}
cf. \cite{WVN}, which leads to a positive and a negative branch of the 
structure function
$g_1^{Q\overline{Q}}(x,Q^2)$ in leading order for a positive definite 
polarized gluon density $\Delta G(x,Q^2)$, see Figure~2a. In this order 
also the first moment of $g_1^{Q\overline{Q}}(x,Q^2)$ vanishes, since  
the r.h.s. of (\ref{eqg1}) is a Mellin--convolution of two functions 
with the support of $C_{g_1}^{Q\overline{Q}}(z,M_Q^2,Q^2)$ being 
$z~\epsilon~[0,1/a]$. Eq.~(\ref{fmom}) also holds for the gluonic
contribution to 
$C_{g_1}^{Q\overline{Q}}(z,M_Q^2,Q^2)$ in the asymptotic limit $Q^2 \gg 
M_Q^2$ in NLO~\cite{BVN}. For quarkonic initial states this relation
does not hold, see~\cite{qm}. 

The choice of the collinear factorization leads to difficulties in 
deriving the correct twist--2 terms in the case of structure
functions which 
contain also twist--3 contributions in the limit of vanishing mass 
scales, as $g_2(x,Q^2)$ and $g_3(x,Q^2)$ in the case of electro--weak 
currents. This was extensively studied in the past~\cite{BK1,BK2,BT} 
using the light--cone expansion and comparing the results to those being 
obtained in parton--model approaches\footnote{For a review of earlier 
results see the comparison given in Ref.~\cite{BK2}.}. In 
Refs.~[5--7] 
it was shown, that for fermionic contributions 
to the 
polarized structure functions the well--known results being obtained in 
the light--cone expansion, see e.g.~\cite{BK2}, can be obtained if one 
refers to the covariant parton model, cf.~\cite{CVPAR}. This is due to 
the fact that the kinematic assumption (\ref{eqSPI}) which neglects all 
parton momenta in the transverse direction is in conflict with the fact 
that the nucleon spin vector has both transverse and longitudinal
components.
This may have impact on predictions of relations between the moments of 
polarized structure functions, the Burkhardt--Cottingham sum 
rule~[22--24]
or integral relations between these functions.
In the following we 
will therefore refer to a general orientation of the parton spin vector 
{\sf and} use the covariant parton model instead of the collinear 
approach.

\section{The general Case}

\vspace{1mm}
\noindent
We now consider the polarization asymmetry of the hadronic tensor
$W_{\mu\nu}^{(A)}$ for a general nucleon spin vector $S_\mu$ and general 
gluon spin vector $s_\mu$, respectively. The nucleon
spin vector obeys~:
\begin{eqnarray}
\label{steq}
S_\mu P^\mu &=& 0,~~~S^\mu = S^\mu_\parallel + S^\mu_\perp\nonumber\\
S^\mu_\parallel &=& (0;0,0,M) \nonumber\\
S^\mu_\perp     &=& M (0;\cos\chi,\sin\chi,0)~.
\end{eqnarray}
The latter two relations hold in the nucleon rest frame.

The polarization asymmetry of the hadronic tensor reads
\begin{eqnarray}
\label{eqWA}
W_{\mu\nu}^{(A)}(q,p,S) &=& \int d^4k \left[f_+(p,k,S) - 
f_-(p,k,S)\right]
{w}_{\mu\nu}^{(A)}(p,q,k,s)~.
\end{eqnarray}
The nucleon spin $S$ is assumed to enter (\ref{eqWA}) linearly as
usually the case in single photon--fermion interactions~\cite{PAUDI}.
Here, $k$ denotes the gluon 4--momentum. 
The
gluon distribution functions $f_\pm(p,k,S)$ refer to opposite proton
spin directions and are supposed to be twist--2 parton distribution
functions in the present paper. The sub--system hadronic tensor asymmetry
${w}_{\mu\nu}^{(A)}(p,q,k,s)$ depends in addition on the virtual 
photon
momentum exchanged, $q$, and the gluonic momentum-- and spin vectors.
$s_\mu$ obeys
\begin{eqnarray}
s_\mu = \frac{p.k}{\sqrt{(p.k)^2 k^2 - M^2 k^4}}
\left[k_\mu - \frac{k^2}{p.k} p_\mu \right]~,
\end{eqnarray}
with $s.k = 0,~k.k = -k^2$. We consider general values of the gluon
virtuality $k^2$,
which is assumed to be sufficiently damped by the difference of the
distribution functions $f_\pm(p,k,S)$ as $k^2 \rightarrow \infty, 0$ in
order to keep (\ref{eqWA}) a well defined relation. Let us denote
\begin{eqnarray}
\sqrt{(p.k)^2 k^2 - M^2 k^4 } = N~.
\end{eqnarray}
The following relations are obtained~:
\begin{eqnarray}
S.s &=& \frac{1}{N}~p.k~S.k \nonumber \\
\frac{S.q}{q.k} &=& \frac{1}{N}~p.k \left( 1 - 
\frac{k^2~p.q}{p.k~q.k}\right)~.
\end{eqnarray}
We define now
\begin{eqnarray}
\Delta f = \frac{M~p.k}{N} \left(f_+ - f_-\right) \equiv
\frac{S.k}{M^2} \widetilde{f}(p^2, p.k, k^2)
\end{eqnarray}
and construct
the sub--system hadronic tensor $w_{\mu\nu}^{(A)}(p,q,k,s)$
for single photon exchange and general values of the virtuality $k^2$.
The Lorentz structure is determined by the Levi--Civita symbol contracted
with two 4--vectors of the problem.
$w_{\mu\nu}^{(A)}(p,q,k,s)$ obeys the representation
\begin{eqnarray}
w_{\mu\nu}^{(A)}(p,q,k,s) &=& i  \varepsilon_{\mu\nu\alpha\beta}
\frac{M~p.k}{N~q.k}
\Biggl\{q^\alpha
\left[k^\beta - \frac{k^2 p^\beta}{p.k}\right]\left[
\widehat{g}_1 + \widehat{g}_2\right] - q^\alpha 
k^\beta \left[1 - \frac{k^2 p.q}
{p.k q.k} \right]
\widehat{g}_2 \nonumber\\
& &
~~~~~~~~~~~~~~~+
k^\alpha \left[k^\beta - \frac{k^2}{p.k}
p^\beta\right] \widehat{v} \Biggr\}~.
\end{eqnarray}
The hadronic tensor is thus given by
\begin{eqnarray}
\label{eqWA1}
W_{\mu\nu}^{(A)}(q,p,S) &=& \frac{i}{M^2}
\varepsilon_{\mu\nu\alpha\beta}
\int d^4 k \widetilde{f}(p^2, p.k, k^2)
\frac{S.k}{q.k}
\left[q^\alpha
k^\beta \left(\widehat{g}_1 + \frac{k^2 p.q}{q.k p.k} \widehat{g}_2
\right) -  \frac{q^\alpha p^\beta k^2}{p.k} \left( \widehat{g}_1
+\widehat{g}_2\right) \right. \nonumber \\
& &~~~~~~~~~~~~~~~~~~~~~~~~~~~~~~~~~~~~~~~\left.- \frac{k^\alpha p^\beta 
k^2}{p.k}~\widehat{v}
\right]~.
\end{eqnarray}
Here  $\widehat{g}_i = \widehat{g}_i(q^2, q.k, k^2)$ and 
$\widehat{v} = \widehat{v}(q^2, q.k, k^2)$ denote the respective 
sub--system structure functions. 
The function $\widehat{v}$ emerges for $k^2 \neq 0$.

For later use we rewrite (\ref{eqWA1}) as
\begin{eqnarray}
W_{\mu\nu}^{(A)}(q,p,S) &=& \frac{i}{M^2} \varepsilon_{\mu\nu\alpha\beta} 
\int d^4 k~S.k~\Biggl\{
 q^\alpha k^\beta \Biggl[
 \frac{\partial a_1}{\partial p.k}
 \frac{\partial b_1}{\partial q.k}
+ p.q~\frac{\partial a_2}{\partial p.k}
 \frac{\partial b_2}{\partial q.k} \Biggr] \nonumber\\
& &~~~~~~~~~~~~~~~~~~~~~~~+ q^\alpha p^\beta \Biggl[
 \frac{\partial a_3}{\partial p.k}
 \frac{\partial b_3}{\partial q.k}
+\frac{\partial a_4}{\partial p.k}
 \frac{\partial b_4}{\partial q.k} \Biggr] \nonumber \\
& &~~~~~~~~~~~~~~~~~~~~~~~+ k^\alpha p^\beta 
 \frac{\partial a_5}{\partial p.k}
 \frac{\partial b_5}{\partial q.k}
\Biggr\}~,
\end{eqnarray}
where
\begin{eqnarray}
 \frac{\partial a_1}{\partial p.k} &=& \widetilde{f}(p^2,p.k,k^2) \\
 \frac{\partial a_i}{\partial p.k} &=& \frac{k^2}{p.k}~
\widetilde{f}(p^2,p.k,k^2),~~{\rm for}~~i = 2 \ldots 5 \\
 \frac{\partial b_1}{\partial q.k} &=& \frac{1}{q.k} 
~\widehat{g}_1(q^2,q.k,k^2) \\
\frac{\partial b_2}{\partial q.k} &=& \frac{1}{(q.k)^2}
~\widehat{g}_2(q^2,q.k,k^2) \\
\frac{\partial b_3}{\partial q.k} &=& -\frac{1}{q.k}
~\widehat{g}_1(q^2,q.k,k^2) \\
\frac{\partial b_4}{\partial q.k} &=& -\frac{1}{q.k}
~\widehat{g}_2(q^2,q.k,k^2) \\
\frac{\partial b_5}{\partial q.k} &=& -\frac{1}{q.k}
~\widehat{v}(q^2,q.k,k^2)~.
\end{eqnarray}
The functions $a_i = a_i(p^2, p.k, k^2)$ and $b_i = b_i(q^2, q.k,k^2)$ 
will be used  in Eq.~(\ref{eqFi}).

The tensor structure in Eq.~(\ref{eqWA1}) is yet different of that in
(\ref{eqHAD}). To determine the structure functions $g_{1,2}(x,Q^2)$ a
tensor decomposition in terms of the
outer variables $p,q$ and $S$ is performed~:
\begin{eqnarray}
\label{eqWA2}
W_{\mu\nu}^{(A)}(q,p,S) &=& \frac{i}{M^2} \varepsilon_{\mu\nu\alpha\beta}
q^\alpha S_\tau\left[I_1^{\beta\tau} +
                     I_2^{\beta\tau} +
                     p^\beta J_1^{\tau} +
                     p^\beta J_2^{\tau} \right]
+ \frac{i}{M^2} \varepsilon_{\mu\nu\alpha\beta} S_\tau p^\beta 
K^{\alpha\tau}~,
\end{eqnarray}
with
\begin{eqnarray}
\label{eqTEN1}
I_1^{\beta\tau} &=& \int d^4 k 
\frac{k^\beta k^\tau}{q.k} 
~\widetilde{f} \cdot \widehat{g}_1
= A_1 g^{\beta\tau} +
B_1 p^\beta p^\tau + C_1 q^\beta q^\tau + D_1\left(p^\beta q^\tau
+ p^\tau q^\beta\right) \\
I_2^{\beta\tau} &=& \int d^4 k 
\frac{k^\beta k^\tau~k^2~p.q}{(q.k)^2 p.k}
~\widetilde{f} \cdot \widehat{g}_2
 = A_2 g^{\beta\tau} +
B_2 p^\beta p^\tau + C_2 q^\beta q^\tau + D_2\left(p^\beta q^\tau
+ p^\tau q^\beta\right) \\
J_1^{\tau} &=& -\int d^4 k 
\frac{k^2~k^\tau}{p.k~q.k} 
~\widetilde{f} \cdot \widehat{g}_1
= E_1 p^\tau + H_1 q^\tau \\
J_2^{\tau} &=& -\int d^4 k 
\frac{k^2~k^\tau}{p.k~q.k} 
~\widetilde{f} \cdot \widehat{g}_2
= E_2 p^\tau + H_2 q^\tau \\
K^{\alpha\tau} &=& -\int d^4 k 
\frac{k^2~k^\alpha k^\tau}{p.k~q.k} 
~\widetilde{f} \cdot \widehat{v}
= A_v g^{\alpha\tau} +
B_v p^\alpha p^\tau + C_v q^\alpha q^\tau + D_v\left(p^\alpha q^\tau
+  q^\alpha p^\tau\right)~.
\end{eqnarray}
The contributions due to $B_{1,2}, C_{1,2}, E_{1,2}, B_v$ and $D_v$
vanish. $A_v$ has to vanish because of current conservation. Therefore
$W_{\mu\nu}^{(A)}$ is given by
\begin{eqnarray}
\label{eqWA3}
W_{\mu\nu}^{(A)}(q,p,S) &=& \frac{i}{M^2} \varepsilon_{\mu\nu\alpha\beta}
q^\alpha S^\beta \left[ A_1 + A_2 \right]
+ \frac{i}{M^2} \varepsilon_{\mu\nu\alpha\beta} q^\alpha p^\beta
S.q \left[D_1 + D_2 + H_1 + H_2 + C_v\right]~.\nonumber\\
\end{eqnarray}
The coefficients $A_{1,2}, D_{1,2}, H_{1,2}$ and $C_v$ read~:
\begin{eqnarray}
\label{eqCOEF1}
A_1 &=& \int d^4 k \left[ \frac{k^2}{2 (q.k)} + \frac{q^2 (p.k)^2}{2 
(p.q)^2 (q.k)} - \frac{(p.k)}{(p.q)} \right] \widetilde{f} 
\cdot \widehat{g}_1 
\\
A_2 &=& \int d^4 k \left[ \frac{k^4 (p.q)}{2 (q.k)^2 (p.k)} + \frac{q^2 
k^2 (p.k)}{2 (p.q) (q.k)^2}
 - \frac{k^2}{(q.k)} \right] \widetilde{f} \cdot \widehat{g}_2 \\
D_1 &=& \int d^4 k \left[- \frac{k^2}{2 (p.q) (q.k)} - \frac{3 q^2
(p.k)^2}{2 (p.q)^3 (q.k)}
 + \frac{2 (p.k)}{(p.q)^2} \right] \widetilde{f} \cdot \widehat{g}_1 \\
D_2 &=& \int d^4 k \left[- \frac{k^4}{2 (p.k) (q.k)^2} - \frac{3 q^2 k^2
(p.k)}{2 (p.q)^2 (q.k)^2}  
 + \frac{2 k^2}{(p.q) (q.k)} \right] \widetilde{f} \cdot \widehat{g}_2 \\
H_1 &=& - \int d^4 k  \frac{k^2}{(p.q) (q.k)}  
\widetilde{f} \cdot \widehat{g}_1 \\
H_2 &=& - \int d^4 k  \frac{k^2}{(p.q) (q.k)}  
\widetilde{f} \cdot \widehat{g}_2 \\
\label{eqCOEF2}
C_v &=& - \int d^4 k  \frac{k^2 (p.k)}{(p.q)^2 (q.k)}
\widetilde{f} \cdot \widehat{v}~. 
\end{eqnarray}

We finally obtain the following representations for 
$g_1(x,Q^2)$ and $g_2(x,Q^2)$~:
\begin{eqnarray}
\label{eqg1A}
g_1(x,Q^2) + g_2(x,Q^2) &=& \frac{p.q}{M^2} [A_1 + A_2] \nonumber\\
          &=& \int d^4 k~\frac{p.q}{M^2} \Biggl\{\Biggl[ \frac{k^2}{2 q.k}
+ \frac{q^2 
(p.k)^2}{2 q.k~(p.q)^2} - \frac{p.k}{p.q} \Biggr]~\widetilde{f} \cdot 
\widehat{g}_1 \nonumber\\ 
& &~~~~~~~+  \Biggl[\frac{k^4~p.q}{2~(q.k)^2
p.k}+\frac{q^2~k^2~p.k}{2 p.q 
(q.k)^2} - \frac{k^2}{q.k} \Biggr]~\widetilde{f} \cdot 
\widehat{g}_2\Biggr\}\nonumber\\
&=& \int d^4 k \Biggl[ \frac{q^2~(p.k)^2}{2 q.k~p.q} - 
p.k \Biggr]~\frac{\widetilde{f} \cdot \widehat{g}_1}{M^2}
+ \int d^4 k \left(\frac{k^2}{Q^2}\right)~\Phi_1(k,p,q)~,
 \\
\label{eqg2A}
g_1(x,Q^2)    &=& \frac{p.q}{M^2} [A_1 + A_2] + \frac{(p.q)^2}{M^2} [D_1 +
D_2 +H_1 + H_2 + C_v] 
\nonumber\\
          &=& \int d^4 k~\frac{p.q}{M^2} \Biggl\{
\Biggl[ -\frac{k^2}{q.k}  
- \frac{q^2
(p.k)^2}{q.k~(p.q)^2} + \frac{p.k}{p.q} \Biggr]~\widetilde{f} \cdot    
\widehat{g}_1 \nonumber\\
& &~~~~~~~~- \Biggl[\frac{q^2 k^2~p.k}{(q.k)^2~p.q}\Biggr]~\widetilde{f} 
\cdot 
\widehat{g}_2 - \Biggl[\frac{p.k~k^2}{q.k~p.q}\Biggr]~\widetilde{f} \cdot
\widehat{v}
\Biggr\} \nonumber\\
&=& \int d^4 k \Biggl[ -\frac{q^2~(p.k)^2}{q.k~p.q} 
+ p.k \Biggr]~\frac{\widetilde{f} \cdot \widehat{g}_1}{M^2}
+ \int d^4 k \left(\frac{k^2}{Q^2}\right)~\Phi_2(k,p,q)~.
\end{eqnarray}
Here the functions $\Phi_{1,2}(k,p,q)$ are finite as $k^2 \rightarrow 0$.

\section{Representation of the structure functions in terms of a 
generating functional}

\vspace{1mm}\noindent
Equivalently to the representation of the polarized structure functions
in the previous section one may represent them referring to a generating 
functional\footnote{This case was studied for polarized light quarks in 
the presence of an on--shell condition in Ref.~\cite{JRR} before.}.
The hadronic tensor can be represented by functions of the form
\begin{equation}
\label{eqFi}
F_i(p^2,p.q,q^2) = \int d^4 k~a_i(p^2,p.k.k^2)~b_i(q^2,q.k,k^2)~,
\end{equation}
where the functions $a_i$ represent the parts depending on $p$ and $k$ and
$b_i$ have a dependence on $q$ and $k$ only. As shown in the foregoing 
this separation is possible accounting for other factors, which do not
depend on $k$. Here $\widetilde{f}$ belongs to the former and
$\widehat{g}_1$, $\widehat{g}_2$, and $\widehat{v}$ to the latter besides
of scalar products.

The partial derivatives of $F$ by $q$ and $p$ can be represented by 
partial derivatives of the functions $a$ and $b$ as follows~:
\begin{eqnarray}
\frac{\partial F}{\partial q^\sigma} &=& p_\sigma \frac{\partial 
F}{\partial p.q} + 2 q_\sigma \frac{\partial F}{\partial q^2}
\nonumber\\
&=& \int d^4k~~a~\cdot~\left[k_\sigma \frac{\partial b}{\partial q.k}
+ 2 q_\sigma \frac{\partial b}{\partial q^2}\right]\\
\frac{\partial F}{\partial p^\sigma} &=& q_\sigma \frac{\partial
F}{\partial p.q} + 2 p_\sigma \frac{\partial F}{\partial p^2}\nonumber\\
&=& \int d^4k~~b~\cdot~\left[k_\sigma \frac{\partial a}{\partial p.k}
+ 2 p_\sigma \frac{\partial a}{\partial p^2}\right]\\
\frac{\partial^2 F}{\partial p^\lambda \partial q^\sigma} &=& 
g_{\lambda\sigma} \frac{\partial F}{\partial p.q} 
+ 2 p_\sigma p_\lambda \frac{\partial^2 F}{\partial p^2 \partial p.q}
+ 2 q_\sigma q_\lambda \frac{\partial^2 F}{\partial q^2 \partial p.q}
+   p_\sigma q_\lambda \frac{\partial^2 F}{\partial (p.q)^2} 
+ 4 q_\sigma p_\lambda \frac{\partial^2 F}{\partial q^2 \partial p^2}  
\nonumber\\
&=& \int d^4 k \Biggl[
k_\lambda k_\sigma \frac{\partial a}{\partial p.k} \frac{\partial 
b}{\partial q.k} + 2 k_\lambda q_\sigma \frac{\partial a}{\partial p.k}
\frac{\partial b}{\partial q^2} 
+2 p_\lambda k_\sigma \frac{\partial a}{\partial p^2} \frac{\partial 
b}{\partial q.k} 
+4 p_\lambda q_\sigma \frac{\partial a}{\partial p^2} \frac{\partial 
b}{\partial q^2} \Biggr]~.
\end{eqnarray}
The following structures contribute to the polarized part of the hadronic 
tensor~:
\begin{eqnarray}
\label{eqEPS1}
\varepsilon_{\mu\nu\alpha\sigma} q^\alpha S_\lambda \int d^4 k
~k^\lambda k^\sigma \frac{\partial a}{\partial p.k} \frac{\partial 
b}{\partial q.k} 
&=& \varepsilon_{\mu\nu\alpha\beta} q^\alpha\left[S^\beta \frac{\partial 
F}{\partial p.q} +S.q p^\beta \frac{\partial^2 F}{\partial 
(p.q)^2}\right]\\
\varepsilon_{\mu\nu\alpha\sigma} p^\alpha S_\lambda \int d^4 k
~k^\lambda k^\sigma \frac{\partial a}{\partial p.k} \frac{\partial
b}{\partial q.k}
&=& \varepsilon_{\mu\nu\alpha\sigma} p^\alpha\left[S^\sigma 
\frac{\partial F}{\partial p.q} + 2 S.q  q^\sigma \frac{\partial^2 F}{\partial
(p.q)^2} \right. \nonumber\\ & &~~~~~~~~\left.
- 2 q^\sigma S_\lambda \int d^4 k~k^\lambda 
\frac{\partial a}{\partial p.k} \frac{\partial
b}{\partial q^2}
\right]  \\
\label{eqEPS2}
\varepsilon_{\mu\nu\alpha\sigma} q^\alpha p^\beta S_\lambda \int d^4 k
k^\lambda  \frac{\partial a}{\partial p.k}~\cdot~b 
&=& 
\varepsilon_{\mu\nu\alpha\beta} q^\alpha p^\beta S.q \frac{\partial 
F}{\partial p.q}~,
\end{eqnarray}
which is given by
\begin{eqnarray}
W_{\mu\nu}^{(A)}(q,p,S) &=& \frac{i}{M^2} \varepsilon_{\mu\nu\alpha\beta}
q^\alpha\Biggl\{S^\beta\Biggl[
\frac{\partial F_1}{\partial (p.q)}+p.q
\frac{\partial F_2}{\partial (p.q)}
\Biggr] 
+ S.q~p^\beta 
  \Biggl[
\frac{\partial^2F_1}{\partial (p.q)^2}+p.q
\frac{\partial^2F_2}{\partial (p.q)^2}
 \nonumber\\ & & ~~~~~~~~~~~~~~~
 + \frac{\partial^2(F_3 + F_4)}{\partial (p.q)^2}\Biggr]
  \nonumber\\ & & ~~~~~~~~~~~~~~~
+ 2 S.q~p^\beta \Biggl[
  \frac{\partial^2 F_5}{\partial (p.q)~\partial q^2}
 -  \int d^4 k \frac{S.k}{S.q}~\frac{\partial a_5}{\partial (p.q)}
 \frac{\partial b_5}{\partial q^2}\Biggr]
\Biggr\}~.
\end{eqnarray}  
The comparison with (\ref{eqHAD}) yields the following expressions for the
structure functions~:
\begin{eqnarray}
\label{eqg1B}
g_1(x,Q^2) + g_2(x, Q^2) &=& \frac{p.q}{M^2} \left[
\frac{\partial F_1}{\partial p.q} + p.q \frac{\partial F_2}{\partial p.q} 
\right]~,\\
\label{eqg2B}
g_2(x,Q^2) &=& - \frac{(p.q)^2}{M^2} \Biggl[
\frac{\partial^2 F_1}{\partial (p.q)^2}+p.q 
\frac{\partial^2 F_2}{\partial (p.q)^2}
+ \frac{\partial^2 (F_3+F_4)}{\partial (p.q)^2}
+ 2 \frac{\partial^2 F_5}{\partial p.q \partial q^2} 
 \nonumber\\ & &~~~~- 2 \int d^4 k 
~\frac{S.k}{S.q} \frac{\partial a_5}{\partial p.q}
 \frac{\partial b_5}{\partial q^2} \Biggr]~.
\end{eqnarray}

We now separate the finite contributions to the polarized structure 
functions in the limit $k^2 \rightarrow 0$ from those which vanish. 
We consider
\begin{eqnarray}
\frac{d}{dx} \left\{x\left[g_1(x,Q^2) + g_2(x, Q^2)\right]\right\} &=& 
-\frac{(p.q)^2}{M^2} \left[
\frac{\partial^2 F_1}{\partial (p.q)^2} + p.q \frac{\partial^2
F_2}{\partial 
(p.q)^2} +
\frac{\partial F_2}{\partial (p.q)}
\right]~,
\end{eqnarray}
where $x= -q^2/(2 p.q)$.
Likewise one obtains
\begin{eqnarray}
\label{eqWW2}
-x \frac{d}{dx} \left\{g_1(x,Q^2) + g_2(x, Q^2)\right\} &=&
g_1(x,Q^2)+g_2(x,Q^2) \nonumber\\ & &
+\frac{(p.q)^2}{M^2} \left[
\frac{\partial^2 F_1}{\partial (p.q)^2} + p.q \frac{\partial^2
F_2}{\partial
(p.q)^2} + \frac{\partial F_1}{\partial (p.q)}
\right]~. 
\end{eqnarray}
On the r.h.s. of (\ref{eqWW2}) one may express the structure function
$g_2(x,Q^2)$ inserting (\ref{eqg2B}) which yields
\begin{eqnarray}  
\label{eqWW3} 
-x \frac{d}{dx} \left\{g_1(x,Q^2) + g_2(x, Q^2)\right\} &=&
g_1(x,Q^2) - \Phi(x,Q^2)~, 
\end{eqnarray}
with
\begin{eqnarray}
\Phi(x,Q^2) &=&
\frac{(p.q)^2}{M^2} \left[  
\frac{\partial (F_3+F_4-F_2)}{\partial (p.q)} + 2 \frac{\partial^2 
F_5}{\partial 
(p.q) \partial q^2} - 2 \int d^4 k \frac{S.k}{S.q} \frac{\partial 
a_5}{\partial 
p.k}
\frac{\partial b_5}{\partial q^2} 
\right]~.
\end{eqnarray}
Let us investigate the structure of the function $\Phi(x,Q^2)$ more 
closely. To do this we refer to (\ref{eqEPS1}--\ref{eqEPS2}) from which 
follows
\begin{eqnarray}
\label{eqX1}
\frac{\partial F_i}{\partial p.q} &=& \int\ d^4 k \frac{S.k}{S.q}
\frac{\partial a_i}{\partial p.k}~\cdot~b~.
\end{eqnarray}
One notices that
\begin{equation}
\label{eqX2}
\frac{\partial a_i}{\partial p.k} =  \frac{k^2}{p.k} \cdot 
\widehat{f},~~~{
\rm for}~~~k=2 \ldots 5~.
\end{equation}
The remainder terms contributing to $\Phi(x,Q^2)$ result from 
(\ref{eqEPS2}). Using Eqs.~(\ref{eqX1},\ref{eqX2}) one thus concludes
that $\Phi(x,Q^2)$ obeys the representation
\begin{equation}
\label{eqPHI1}
\Phi(x,Q^2) = \int d^4 k \left(\frac{k^2}{Q^2}\right)~ 
\phi(p.k,q.k,p^2,q^2,k^2)~, 
\end{equation} 
where $\phi(p.k,q.k,p^2,q^2,k^2)$ is finite for $k^2 \rightarrow 0$.

\section{The relation between {\boldmath $g_1(x,Q^2)$} and {\boldmath 
$g_2(x,Q^2)$}} 

\vspace{1mm}
\noindent   

In (\ref{eqg1A}, \ref{eqg2A}, \ref{eqg1B}, \ref{eqg2B}) the virtuality of 
the gluon field  is revealed by power corrections in $(k^2/Q^2)^l$.   
These functions are not yet projected onto the twist--2 
contributions.\footnote{Plain consideration of off--shell contributions 
in $k^2$ to hadronic structure functions may introduce unphysical 
contributions. One example are wrong target mass corrections as noted in 
\cite{TM1} a long time ago.} Any kind of partonic approach is only valid
{\sf if} the partonic virtualities $k^2$ obey 
\begin{equation}
\label{eqDY}
|k^2| \ll Q^2~.   
\end{equation}
This is an analogous condition to the requirement that the parton 
lifetime has to be much larger than the interaction time in the deeply 
inelastic scattering process for all associated infinite momentum 
frames,~\cite{DY}.
To project out the twist--2 contributions 
we refer to the collinear basis, cf.~\cite{EFP}, and expand these 
functions into a Taylor series in $k^\mu$ at $k^\mu = z p^\mu$. Here
$p.p =0$ and the lowest twist contribution is obtained in setting $k^2 
\rightarrow 0$ in (\ref{eqg1A}, \ref{eqg2A}, \ref{eqg1B}, \ref{eqg2B}). 
Note, however, that the corresponding expressions may contribute at 
higher twist due to the associated derivatives in $k^\mu$.
We denote the twist--2 contributions to the structure functions
$g_i(x,Q^2)$ by $g_i^{\rm II}(x,Q^2)$. 

Whereas the relation between the structure functions $g_1^{\rm II}$ and
$g_2^{\rm II}$ is not easily seen from Eqs.~(\ref{eqg1A}, \ref{eqg2A})
\footnote{For this the explicit dependence on $(p.q)$ had to be known
for Eq.~(\ref{eqg1A}).}, it can be directly obtained from (\ref{eqWW3}),
\begin{equation}
\label{eqWW4}
- x\frac{d}{dx}\left[g_1^{\rm II}(x,Q^2)+g_2^{\rm II}(x,Q^2)\right] = 
g_1^{\rm II}(x,Q^2)~,   
\end{equation}
the differential form of the  Wandzura--Wilczek relation,
see~\cite{WW,BT}. The integral form can be obtained from
(\ref{eqWW4}) using the condition
\begin{equation}
\label{eqLIM}
\lim_{x \rightarrow 1} g_i^{\rm II}(x,Q^2) = 0
\end{equation}
which yields
\begin{equation}
\label{eqWW}
g_2^{\rm II}(x,Q^2) = - g_1^{\rm II}(x,Q^2) + \int_x^1 \frac{dy}{y}
g_1^{\rm II}(y,Q^2)~.
\end{equation}
At the level of twist--2 all structure functions depend on the same
non--perturbative function $\widetilde{f}$ {\sf and}
sub--system structure function $\widehat{g}_1^{\sf par}$. Since the
Wandzura--Wilczek
relation holds as well for quarkonic initial states~\cite{RR,BK1}
all higher order radiative corrections can be absorbed into 
$\widehat{g}_1^{\sf par}$, respectively, with ${\sf par}~=g,~q_i$
denoting the quark species and the gluon, and holds thus in all orders. 

The validity of the Wandzura--Wilczek relation also in the case of gluonic
operator matrix elements is in accordance with the observation of the
general nature of this relation connecting vector operator
matrix--elements with the associated scalar operator matrix--elements on the
light cone, which was shown for fermionic fields in \cite{BGR,BR1}.

To compare the heavy flavor contributions to $g_{1,2}(x,Q^2)$ to the usual
parameterizations of these structure functions numerically we show the light
flavor contributions to $xg_{1,2}(x,Q^2)$ in Figures~1a,b at leading order. For 
the parton distributions we refer to the parameterization~\cite{BB}. Other
recent parameterizations~\cite{GRSV,AAC} are well in accordance with \cite{BB}
within the current experimental errors. The polarized gluon and sea--quark
distributions have still a rather large uncertainty which can only be lowered 
by more precise data from future experiments. The present
parameterizations 
were derived assuming three light quarks through a fit to the data ignoring 
any heavy flavor contribution. Part of the present gluon--distribution is thus
corresponding to the contribution due to heavy flavors and one cannot add the
distributions in Figures~1a,2a or Figures~1b,2b in a simple way.

As Figure~2a shows the effect of $xg_1^{c\overline{c}}(x,Q^2)$ is small at low
scales $Q^2$ but rises rapidly with $Q^2$ and should be taken into account in
future refined QCD analyses. Due to the larger quark mass and 
charge--suppression the $b$--quark contribution to $xg_1(x,Q^2)$ is  smaller
than that due to charm--quarks. While the distributions are positive for $x >
5 \cdot 10^{-3}$ they change sign for lower values of $x$. In the present
example the polarized gluon distribution is positive. The Wilson coefficient 
(\ref{eqWILS}) changes sign, since
\begin{eqnarray}
\lim_{z \rightarrow 0} C_{g_1}^{Q\overline{Q}}(z,Q^2) &=& \frac{1}{2}\left[
3 - \ln\left(\frac{Q^2}{z M_Q^2}\right) \right]~,
\\
\lim_{\beta \rightarrow 0} C_{g_1}^{Q\overline{Q}}(z,Q^2) &=& \beta > 0,
\end{eqnarray}
and vanishes at threshold $z = Q^2/(1+4M^2_Q)$. Furthermore 
Eq.~(\ref{fmom}) holds. The oscillating behaviour of the Wilson
coefficient implies lower relative heavy flavor contributions than in
the unpolarized case.

Figure~1b shows $xg_2(x,Q^2)$ for the light flavors due to the
Wandzura--Wilczek relation. $xg_2(x,Q^2)$ is positive for small $x$ values up to
$x \sim 10^{-1}$ and changes sign then. The integral over the {\sf positive}
function $xg_1(x,Q^2)$ is such larger than the subtraction term $xg_1(x,Q^2)$
in the former region while the subtraction term dominates in the latter region.
The function has to have a positive and a negative branch since the
Wandzura--Wilczek relation formally covers the Burkhardt--Cottingham 
relation\footnote{Note that the Wandzura--Wilczek relation is the
analytic continuation from the {\sf positive} moments, cf. e.g.
\cite{BK2,BT} in the local light cone expansion, where the 0th moment,
which corresponds to the Burkhardt--Cottingham sum rule, does not 
contribute.},
\begin{eqnarray}
\label{eqBC}
\int_0^1 dx g_2(x,Q^2) = 0~.
\end{eqnarray}
Due to the additivity of twist--2 structure functions w.r.t. their parton
contents the relation has to hold for each flavor separately.

Similar conclusions can be drawn for $xg_2^{Q\overline{Q}}(x,Q^2)$ shown 
in
Figure~2b. $xg_2^{Q\overline{Q}}(x,Q^2)$ is widely positive for smaller
values of $x \lsim 2 \cdot 10^{-2}$
and negative for larger $x$. Again we would like to stress that the
present numerical results on $xg_{1,2}^{Q\overline{Q}}(x,Q^2)$ are very
sensitive on the polarized gluon distribution.

\section{Conclusions} 

\vspace{1mm}
\noindent
We calculated the heavy flavor contribution to $g_2(x,Q^2)$ at leading order in
the strong coupling constant using the covariant parton model for finite values
of the gluon virtuality $k^2$. The representation of the polarized structure
functions $g_{1,2}^{Q\overline{Q}}(x,Q^2)$ can be obtained applying a tensor
decomposition. Furthermore, a generating functional in which the
$k$--dependent
parton densities and coefficient functions are connected can be used to obtain a
representation of the structure functions from which their possible relation
can be derived. The twist--2 contributions are obtained in the limit $k^2 \ll
Q^2$. The functions $g_{1,2}^{Q\overline{Q}}(x,Q^2)$ obey the Wandzura--Wilczek
relation for a gluon induced process similar to earlier findings in fermion
induced processes~[5--10,12,13,15--18]. The absolute value of the
relative
numerical effect on both structure functions due to the LO heavy flavor
contributions is about the same and may reach values of up to 5--10\% in
some kinematic ranges.~\footnote{Similar size effects have been reported 
in \cite{BMSN} for $g_1^{Q\overline{Q}}$ before.} 
To make future QCD analyses of even more precise
experimental
data consistent at the level of the twist--2 contributions the heavy flavor
distributions have to be taken into account.



\newpage
\begin{center}

\mbox{\epsfig{file=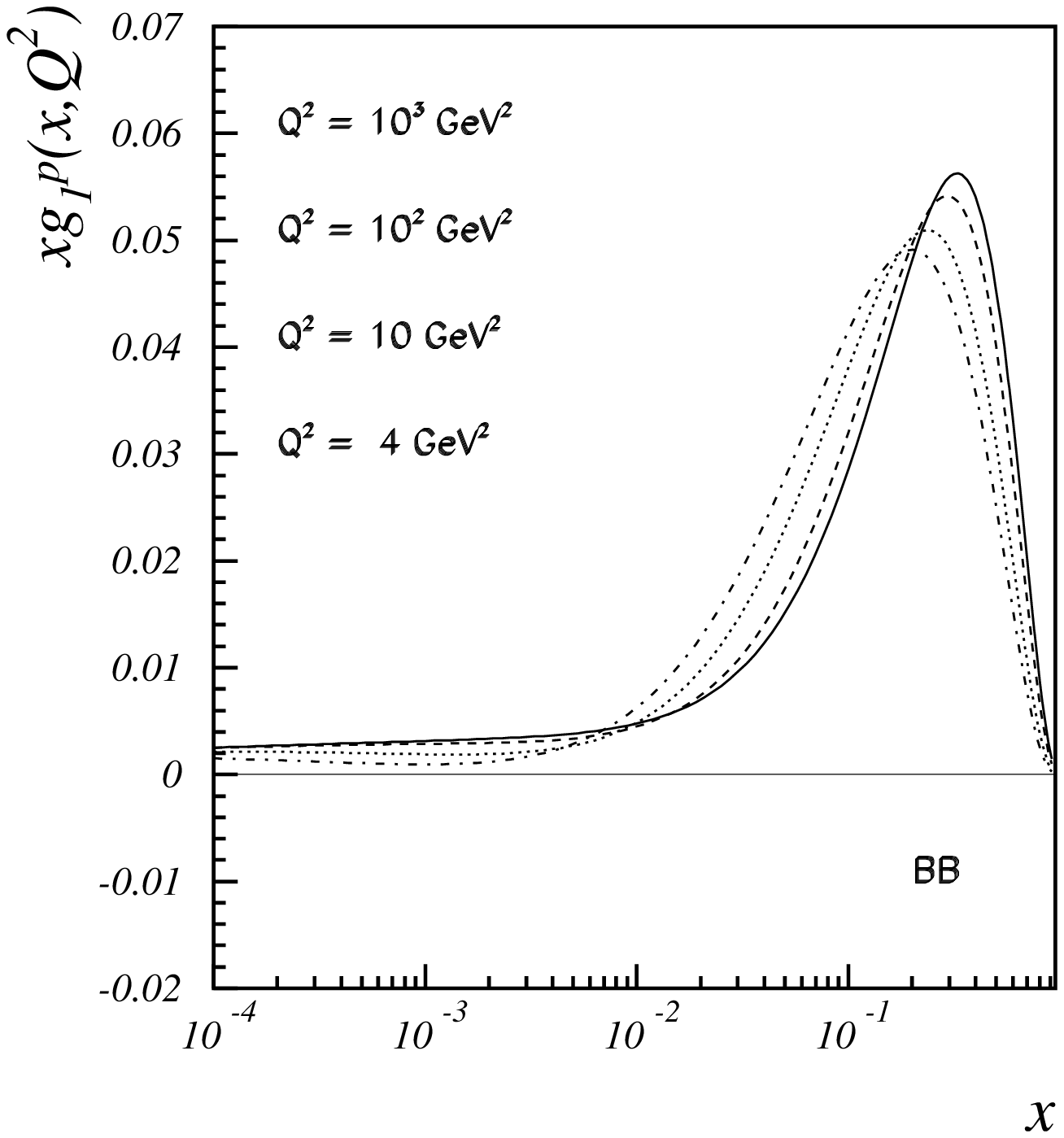,height=16cm,width=16cm}}

\vspace{2mm}
\noindent
\small
\end{center}
{\sf
Figure~1a:~The light flavor contributions to the polarized structure
function $xg_1^p(x,Q^2)$ in leading order (parameterization {\tt ISET=1}
of \cite{BB}, $\Lambda_{\rm QCD} = 203 \MeV$) as a function of $x$ and
$Q^2$. Full line: $Q^2 = 4 \GeV^2$; dashed line: $Q^2 = 10 \GeV^2$; dotted
line: $Q^2 = 100 \GeV^2$; dash--dotted line: $Q^2 = 1000 \GeV^2$.} 
\normalsize
\begin{center}

\mbox{\epsfig{file=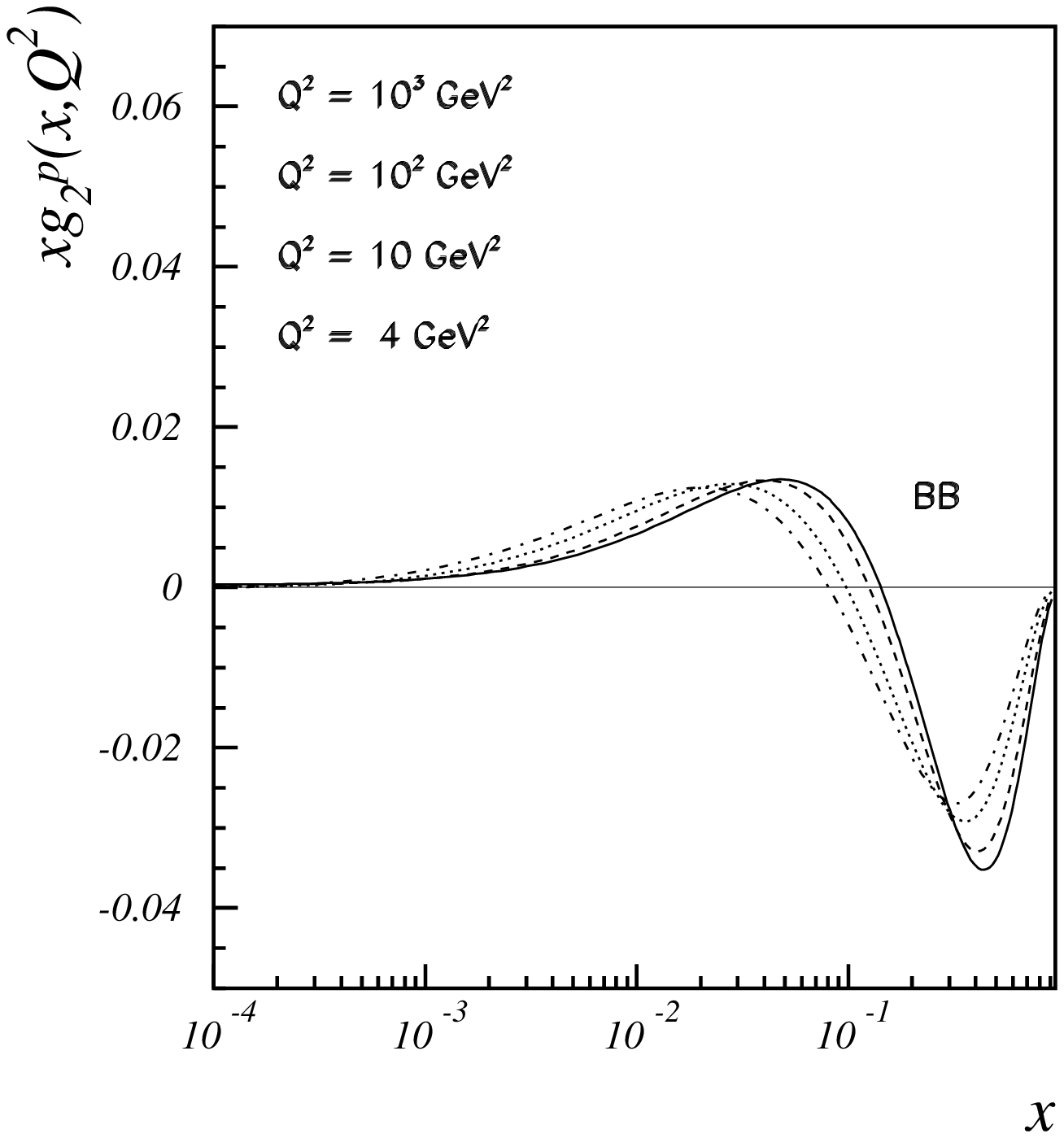,height=16cm,width=16cm}}

\vspace{2mm}
\noindent
\small
\end{center}
{\sf
Figure~1b:~The light flavor contributions to the polarized structure
function $xg_2^p(x,Q^2)$. All conditions are the same as in Figure~1a.
\normalsize
\newpage

\vspace*{-1cm}
\begin{center}

\mbox{\epsfig{file=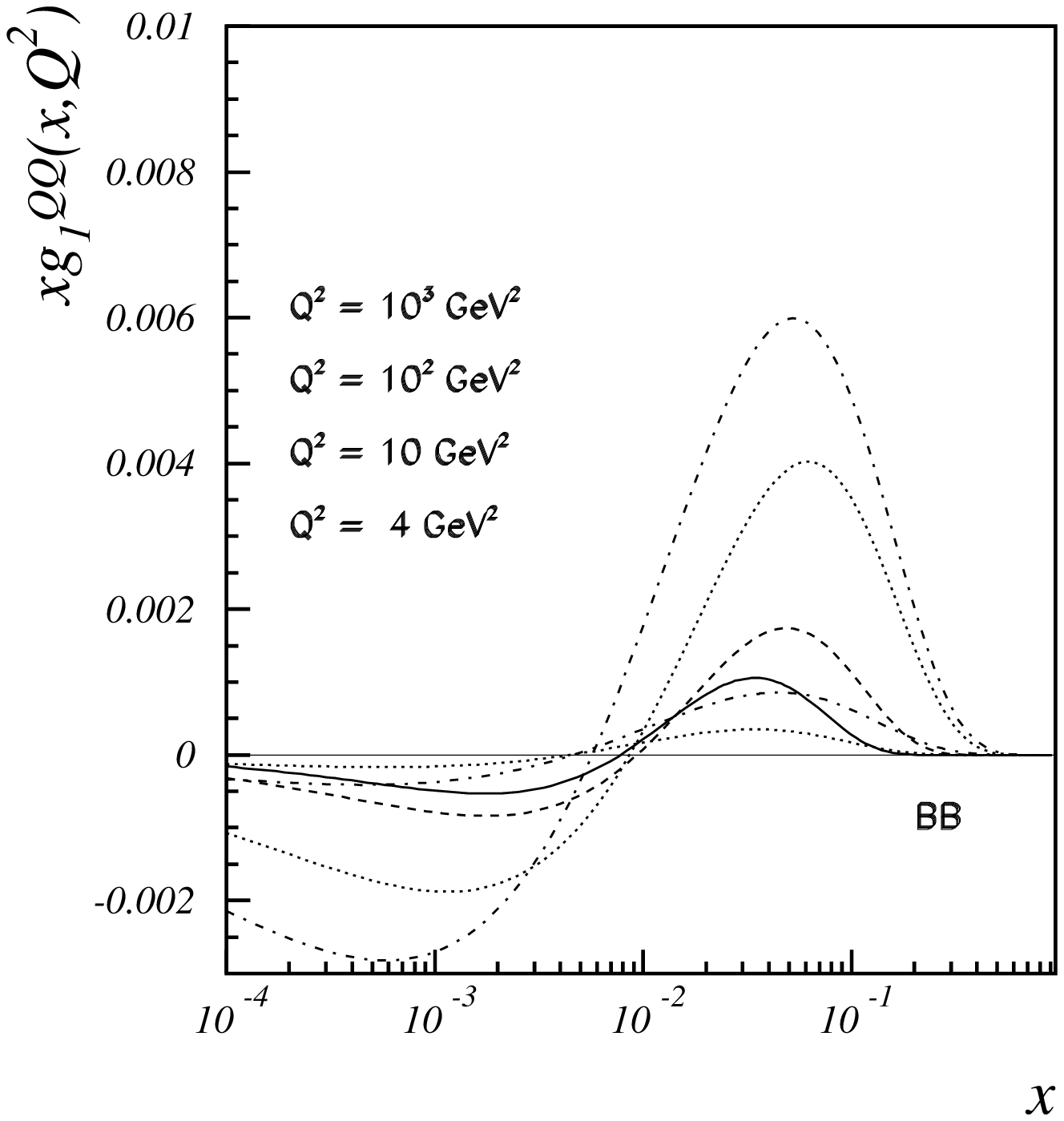,height=16cm,width=16cm}}

\vspace{2mm}
\noindent
\small
\end{center}
{\sf
Figure~2a:~Heavy flavor contributions to the polarized structure
function $xg_1(x,Q^2)$ in leading order (parameterization {\tt ISET=1}
of \cite{BB}, $\Lambda_{\rm QCD} = 203 \MeV$) as a function of $x$ and
$Q^2$. Upper lines: $g_1^{c\overline{c}}(x,Q^2)$ for $m_c = 1.5 \GeV$.
Full line: $Q^2 = 4 \GeV^2$; dashed line: $Q^2 = 10 \GeV^2$; dotted
line: $Q^2 = 100 \GeV^2$; dash--dotted line: $Q^2 = 1000 \GeV^2$.
Lower lines: $g_2^{c\overline{c}}(x,Q^2)$ for $m_b = 4.3 \GeV$.
Dotted line: $Q^2 = 100 \GeV^2$; dash--dotted line: $Q^2 = 1000 \GeV^2$.}   
Figure~2b:~Heavy flavor contributions to the polarized
structure 
function $xg_2^p(x,Q^2)$.
\normalsize

\vspace*{-1cm}
\begin{center}

\mbox{\epsfig{file=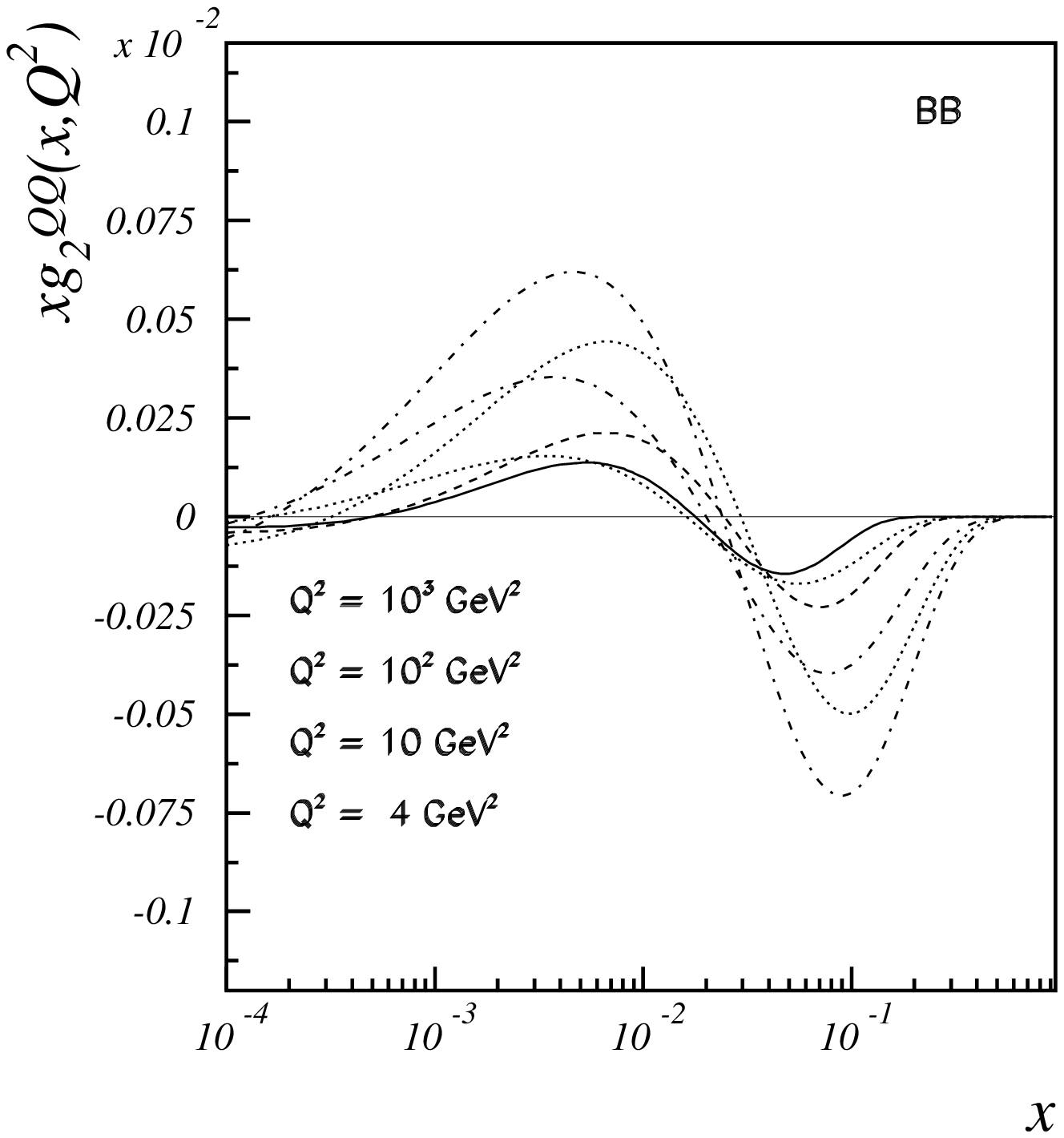,height=16cm,width=16cm}}

\vspace{2mm}
\noindent
\small
\end{center}
{\sf  
Figure~2b:~Heavy flavor contributions to the polarized structure
function $xg_2^p(x,Q^2)$.
All conditions are the same as in Figure~2a.
\normalsize


\begin{thebibliography}{99}
%
\bibitem{UNPLA}
J. Breitweg et al., ZEUS collab., Eur. Phys. J. {\bf C12} (2000) 35;\\
C. Adloff et al., H1 collab., Phys. Lett. {\bf B528} (2002) 199.
%
\bibitem{UNP}
E. Laenen, S. Riemersma, J. Smith, and W.L. van Neerven, Phys. Lett. {\bf 
B291} (1992) 325; Nucl. Phys. {\bf B392} (1993) 162, 229;\\
S. Riemersma, J. Smith, and W.L. van Neerven, Phys. Lett. {\bf B347} 
(1995) 143;\\
T. Gottschalk, Phys. Rev. {\bf D23} (1981)  56;\\
G. Kramer and B. Lampe, Z. Phys. {\bf C54} (1992) 139; E: Eur. Phys. J.
{\bf C17} (2000) 371.
%
\bibitem{BVN}
M. Buza and W.L. van Neerven, Nucl. Phys. {\bf B500} (1997) 420. 
%
\bibitem{POLLO}
A.D. Watson, Z. Phys. {\bf C12} (1982) 123;\\
W. Vogelsang, Z. Phys. {\bf C50} (1991) 275.
%
\bibitem{JRR}
J.D. Jackson, G.G. Ross, and R.G. Roberts, Phys. Lett. {\bf B226} (1989) 
159.
%
\bibitem{RR}  
R.G. Roberts and G.G. Ross, Phys. Lett. {\bf B373} (1996) 235.
%
\bibitem{BK1}
J. Bl\"umlein and N. Kochelev, Phys. Lett. {\bf B381} (1996) 296. 
%
\bibitem{BK2}
J. Bl\"umlein and N. Kochelev, Nucl. Phys. {\bf B498} (1997) 285. 
%
\bibitem{WW}
S. Wandzura and F. Wilczek, Phys. Lett. {\bf B72} (1977) 195.
%
\bibitem{BT}
J. Bl\"umlein and A. Tkabladze,
Nucl. Phys. {\bf B553} (1999) 427; 
Nucl. Phys. {\bf B} (Proc. Suppl.) {\bf 79} (1999) 541. 
%
\bibitem{GP}
H. Georgi and H.D. Politzer, Phys. Rev. {\bf D14} (1976) 1829.
%
\bibitem{RID}
A. Piccione and G. Ridolfi, Nucl. Phys. {\bf B513} (1998) 301.
%
\bibitem{BGR}
J. Bl\"umlein, B. Geyer and D. Robaschik, Nucl. Phys. {\bf B560} (1999) 283. 
%
\bibitem{CG}
C.G. Callan and D.J. Gross,  Phys. Rev.  Lett. {\bf 22} (1969) 156.  
%
\bibitem{BR1}
J. Bl\"umlein and D. Robaschik, Nucl. Phys. {\bf B581} (2000) 449. 
%
\bibitem{BEGR}
J. Bl\"umlein, J. Eilers, B. Geyer, and D. Robaschik,
Phys. Rev. {\bf D65} (2002) 054029.
%
\bibitem{BR2}
J. Bl\"umlein and D. Robaschik,  Phys. Lett. {\bf B517} (2001) 222; 
Phys. Rev. {\bf D65} (2002) 096002. 
%
\bibitem{GL}
B. Geyer and M. Lazar, Phys. Rev. {\bf D63} (2001) 094003;\\
J. Bl\"umlein, B. Geyer, M. Lazar, and D. Robaschik, Nucl. Phys.
{\bf B} (Proc. Suppl.) {\bf 89} (2000) 155.
%
\bibitem{WVN}
W.L. van Neerven, {\tt hep-ph/9709569}.
%
\bibitem{qm}
J. Bl\"umlein and W.L. van Neerven, Phys. Lett. {\bf B450} (1999) 417. 
%
\bibitem{CVPAR}
C. Nash, Nucl. Phys. {\bf B31} (1971) 419;\\
P.V. Landshoff and J.C. Polkinghorne, Phys. Rep. {\bf C5} (1972) 1.
%
\bibitem{BC}
H. Burkhardt and W.N. Cottingham, Ann. Phys. (New York) {\bf 56} (1970) 
453.
%
\bibitem{MN}
R. Mertig and W.L. van Neerven, Z. Phys. {\bf C60} (1993) 489; E: {\bf C65}
(1995) 360;\\
G. Altarelli, B. Lampe, P. Nason and G. Ridolfi,
Phys. Lett. {\bf  B334} (1994) 187;\\
J. Kodaira, S. Matsuda, T. Uematsu and K. Sasaki, Phys. Lett. {\bf B345} 
(1995) 527. 
%
\bibitem{RN}
V. Ravindran and W.L. van Neerven, Nucl. Phys. {\bf B605} (2001) 517.
%
\bibitem{PAUDI}
W. Pauli, Z. Physik. {\bf 43} (1927) 601;\\
P.A.M. Dirac, Proc. Roy. Soc. (London) {\bf A117} (1928) 610.
%
\bibitem{TM1}
R.K. Ellis, R. Petronzio, and Parisi,Phys. Lett. {\bf B64} (1976) 97; \\
R. Barbieri et al., Phys. Lett. {\bf B64} (1976) 171; Nucl. Phys. {\bf
B117} (1976) 50.
%
\bibitem{DY}
S.D. Drell and T.-M. Yan, Ann. Physics (New York) {\bf 66} (1971) 578.
%
\bibitem{EFP}
R.K. Ellis, W. Furmanski, and R. Petronzio, Nucl. Phys. {\bf B212} (1983) 
29.
%
\bibitem{BB}
J. Bl\"umlein and H. B\"ottcher, Nucl. Phys. {\bf B636} (2002) 225. 
%
\bibitem{GRSV}
M. Gl\"uck, E. Reya, M. Stratmann, and W. Vogelsang, Phys. Rev. {\bf D63}
(2001) 09400.
%
\bibitem{AAC}
Y. Goto et al., AAC collab., Phys. Rev. {\bf D62} (2000) 034017;\\
M. Hirai, {\tt hep-ph/0211190}.  
%
\bibitem{BMSN}
M. Buza, Z. Matiounine, J. Smith, W.L. van Neerven, Nucl. Phys. {\bf
B485} (1997) 420. 
\end{thebibliography}
\end{document}